\documentclass[aps,pre,onecolumn,superscriptaddress]{revtex4}

\usepackage{amssymb}
\usepackage{amsfonts}
\usepackage{rotating}
\usepackage{amsmath}
\usepackage{graphics}
\usepackage{epsfig}
\usepackage[pdftex,usenames]{color}
\usepackage{hyperref}
\usepackage{makecell}

\newcommand{\be}{\begin{equation}}
\newcommand{\ee}{\end{equation}}

\newcommand{\bea}{\begin{eqnarray}}
\newcommand{\eea}{\end{eqnarray}}

\begin{document}

%\draft

\title{Explore with caution: mapping the evolution of scientific interest in Physics}

\author{Alberto Aleta}
\affiliation{Department of Theoretical Physics, University of Zaragoza, Zaragoza 50009, Spain}
\affiliation{Institute for Biocomputation and Physics of Complex Systems (BIFI), University of Zaragoza, 50018, Zaragoza, Spain}

\author{Sandro Meloni}
\affiliation{Institute for Biocomputation and Physics of Complex Systems (BIFI), University of Zaragoza, 50018, Zaragoza, Spain}
\affiliation{IFISC, Institute for Cross-Disciplinary Physics and Complex Systems (CSIC-UIB), 07122 Palma de Mallorca, Spain}

\author{Nicola Perra}
\affiliation{Centre for Business Networks Analysis, University of Greenwich, London, UK}
\affiliation{Institute for Scientific Interchange, ISI Foundation, Turin, Italy}

\author{Yamir Moreno}
\affiliation{Department of Theoretical Physics, University of Zaragoza, Zaragoza 50009, Spain}
\affiliation{Institute for Biocomputation and Physics of Complex Systems (BIFI), University of Zaragoza, 50018, Zaragoza, Spain}
\affiliation{Institute for Scientific Interchange, ISI Foundation, Turin, Italy}

\date{\today} \widetext

\begin{abstract}
In the book {\it The Essential Tension}~\cite{kuhn1979essential} Thomas Kuhn described the conflict between tradition and innovation in scientific research --i.e., the desire to explore new promising areas, counterposed to the need to capitalize on the work done in the past. While it is true that along their careers many scientists probably felt this tension, only few works have tried to quantify it. 
Here, we address this question by analyzing a large-scale dataset, containing all the papers published by the American Physical Society (APS) in more than $25$ years, which allows for a better understanding of scientists' careers evolution in Physics. We employ the Physics and Astronomy Classification Scheme (PACS) present in each paper to map the scientific interests of $181,397$ authors and their evolution along the years. Our results indeed confirm the existence of the ``essential tension'' with scientists balancing between exploring the boundaries of their area and exploiting previous work. In particular, we found that although the majority of physicists change the topics of their research, they stay within the same broader area thus exploring with caution new scientific endeavors. Furthermore, we quantify the flows of authors moving between different subfields and pinpoint which areas are more likely to attract or donate researchers to the other ones.
Overall, our results depict a very distinctive portrait of the evolution of research interests in Physics and can help in designing specific policies for the future.   
\end{abstract}

%\pacs{xx,xx}
\maketitle

Take a second and think of the main topic of your latest publication. Is it the same of the paper you are currently working on? If you are in the academic business, chances are that the answer to this question is yes. In the case, instead, the answer is no, how far the two topics are?  What does far, in this context, even mean?

It is long been acknowledged that researchers are constantly pulled by two opposite forces: the exploration of new directions and the exploitations of an established research agenda~\cite{kuhn1979essential,foster2015tradition,march1991exploration,rzhetsky2015choosing,jia2017quantifying}. The former can lead to ground breaking results, radical new knowledge, acclaim and success, but it is a risky strategy often linked to failure, decrease in productivity and challenges in pushing forward ideas in new academic circles~\cite{merton1957priorities,kuhn1970structure}. The latter, instead, is a conservative strategy associated to high chances of steady publications outputs, fair visibility, but it is typically linked to incremental and low-impact as well as low originality outputs~\cite{foster2015tradition}. Thomas Kuhn eloquently defined this conflictual situation as ``the essential tension'' between risky and conservative strategies~\cite{kuhn1979essential}. In specific fields such tension has been defined as the perennial fight between conformity and dissent (philosophy of science~\cite{polanyi1969knowing}), succession and subversion (sociology of science~\cite{whitley2000intellectual}) or refinement and risk taking (innovation~\cite{march1991exploration}). 

Societal progress, academic success, policies, and funding allocation are the complex outcome of scientists reactions and interactions with this tension. Therefore, it is of crucial importance to quantify and understand how scientific interest, and consequently science, evolves in time. To this end, the digitalisation of publication records is of great help~\cite{fortunato2018science,clauset2017data}. Authors, affiliations, references, text, and various tags of virtually any publication are now digitally collected (also retrospectively) and stored in databases. The access to such data, often limited to specific journals and/or fields, has boosted the number of studies investigating publication/citation patterns of authors~\cite{Hirsch2005,Egghe2006,Hirsch2007,mukherjee2017nearly,deville2014career,petersen2015quantifying,guevara2016research}, papers~\cite{Redner1998,Chen2007}, journals~\cite{Garfield1972,Bergstrom2007}, institutions~\cite{Borner2006,jones2008multi,guevara2016research}, cities~\cite{Bornmann2011}, or countries~\cite{King2004,zhang2013characterizing,guevara2016research}. Arguably, the most popular area of investigation is the development of metrics aimed at ranking scientific outputs at different granularities (from single authors to countries)~\cite{Hirsch2005,waltman2016review,kaur2013universality,sinatra2016quantifying,SARA,wang2013quantifying,radicchi2008universality,fraiberger2018quantifying,liu2018hot,lehmann2008quantitative}. Instead, studies aimed at quantifying or understanding the effects of the ``essential tension'' mentioned above received far less attention. 

Before moving to describe our contribution in this underdeveloped area, we believe that it is important to briefly summarise four recent papers that did focus on such topic and are close to our aims. Foster et al~\cite{foster2015tradition}, studied researchers strategies in the area of biomedical chemistry. Using tools from Network Science, they studied the evolution of knowledge in the field and found that i) despite the growth of the field in time the distribution of strategies remains constant ii) exploration (high-risk strategies) is less prevalent than exploitation (low-risk strategies) iii) exploration is more likely to be ignored, but when it is not, it is linked to high impact and success. Pan et al~\cite{pan2012evolution}, considered the papers published by the American Physical Society (APS) and use tools from Network Science to map the evolution of scientific progress and thus interest in specific topics across time. They built annual networks connecting topics, defined via the Physics and Astronomy Classification Scheme (PACS), if two were listed in the same paper. By studying the properties of such networks they characterised the systemic effects of research strategies of exploration and/or exploitation. They found that i) the statistical features of such networks are quite stationary across time ii) there is an overall increase in connectivity between different fields iii) the unfolding of such increase is hierarchical (closer topics get connected first than far ones) iv) the networks are dominated by topics belonging to subfields of Condense Matter and General Physics, and v) there is an increase in the importance of Interdisciplinary Physics. Jia et al~\cite{jia2017quantifying} also studied the APS dataset focusing on PACS. However, they considered the evolution of interest between topics in the careers of single authors. They found that the empirical patterns can be explained by an interplay between exploration and exploitation modulated by three factors: heterogeneity, recency, and subject proximity. Very recently, Battiston et al~\cite{battiston2019taking} presented the most comprehensive analysis (to the best of our knowledge) of Physics to the date. Using tools from Network and Data Science, they analysed the Web of Science and reconstructed the career of about $135,000$ physicists by considering $294$ Physics journals and many more interdisciplinary venues. They adopted PACS to classify the topic(s) and thus the field(s) of Physics represented in each publication. By leveraging this dataset they provided the ``census'' of different fields of Physics, studied the movement and transition of physicists between them, studied the role of chaperones, quantified differences between fields (considering frequency of publication, collaboration size, and citations), and studied the recognition (i.e. Nobel prizes) of each area of Physics. Although, the focus of their research was not the tension between exploration and exploitation, their analysis of the transitions between fields highlighted interesting patterns: i) Condensed Matter is the starting field of many physicists that then move to Interdisciplinary, Classical, and General Physics, ii) High Energy and Nuclear Physics tend to ``swap'' scientists that might also move towards Astrophysics, and iii) Plasma and Astrophysics are the fields that ``welcome'' more physicists from different backgrounds.\\

In this context, we study the APS dataset considering the period between $1980$ and $2006$. We use the PACS associated to each paper and investigate the evolution of interest between topics in the careers of scientists. To this end, we first quantify  the tendency towards exploration and exploitation measuring the similarity, in terms of topics, between the production during the first and last year of activity of each author. We then deepen the analysis characterizing the transition patterns between sub-fields. In particular, we build source (first year of activity) - destination (last year of activity) matrices and study the networks flows between them. Finally, we study the transitions between fields as a function of time considering the entire career of each author. Our results depict a peculiar landscape with authors balancing between the desire to explore new topics and the need of exploiting the acquired knowledge. These trends seem also to be stable in the last $30$ years allowing us to highlight the future evolution paths of the distinct areas of Physics. It is important to mention that although our objectives are aligned with the four papers mentioned above, here we develop/adopt different and complementary metrics. Thus, our results contribute to uncover the complex dynamics of scientific production in time focusing on the tension between exploration and exploitation that any researcher likely faces. 

\section{Dataset}

We consider the APS dataset which includes all papers published by the Society from $1893$ to $2009$. As we are interested in the evolution of interest between topics, we use PACS. This classification scheme has been developed since $1970$. The final PACS classification has been released in $2010$ and it has been in use in the APS journals till $2016$, when the APS introduced a new classification scheme called PhySH (Physics Subject Headings) that is substituting PACS. Our raw piece of information is the evolution of interest of each single author measured through the use of PACS. Thus, we need to know which author published which paper. Given that the process of disambiguation of authors names is per se a scientific challenge, we decided to use the dataset outcome of Ref.~\cite{SARA} (we invite the interested reader to the original paper for all the details of the process). Considering the various constrains (both in terms of PACS and authors disambiguation availability) in the following we analysed all the papers published between $1980$ and $2006$. This includes $270,781$ papers, published in $9$ journals, by $181,409$ authors.

The PACS classification scheme is organised as a tree composed by four levels. To better understand its structure let us consider the following PACS number $05.70.Ce$ which indicates papers dealing with ``thermodynamic functions and equations of state".  The first digit ($0$) describes the first level: General Physics. This can be chosen among $10$ (from $0$ to $9$). The first and second digit ($05$) describe the second level: Statistical Physics, Thermodynamics, and Nonlinear Dynamical Systems. There are $68$ ids at depth $2$ in the classification tree in our dataset. The third level is constituted by the first two digits and by the second number ($05.70$), Thermodynamics in this case. At the more granular level we need to add the two letters and get the complete description of the PACS given before. To guide the reader to understand what follows, in Table~\ref{table:1} we report  the ids and names associated to the first level of the classification tree.

\begin{table}[!ht]
\centering

\resizebox{0.8\textwidth}{!}{
\begin{tabular}{@{}ll@{}}
\toprule
\textbf{Id}                        &  \textbf{Description}  \\ \colrule
0 & General Physics \\
1& The Physics of Elementary Particles and Fields\\
2& Nuclear Physics \\
3& Atomic and Molecular Physics \\
4& Electromagnetism, Optics, Acoustics, Heat Transfer, Classical Mechanics, and Fluid Dynamics\\
5 & Physics of Gases, Plasmas, and Electric Discharges\\
6 & Condensed Matter: Structural, Mechanical and Thermal Properties\\
7& Condensed Matter: Electronic Structure, Electrical,Magnetic, and Optical Properties\\
8& Interdisciplinary Physics and Related Areas of Science and Technology\\
9 & Geophysics, Astronomy, and Astrophysics \\
\botrule
\end{tabular}
}
\caption{Description of the first level of the classification scheme}
\label{table:1}
\end{table}

\section{Results}
\begin{figure}
\includegraphics[scale=0.3]{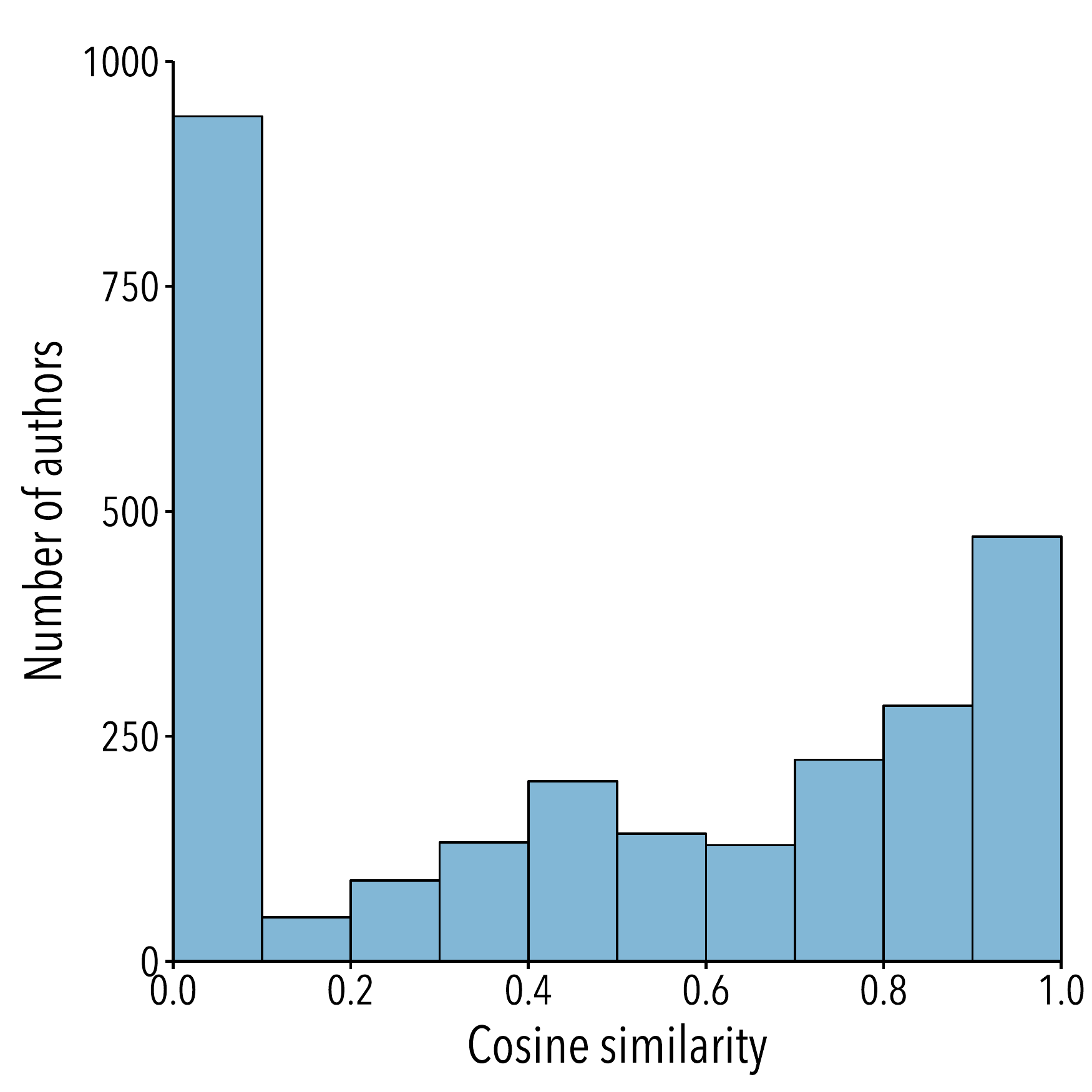}
\caption{Cosine similarity distribution of authors' interest who started their careers (published their first paper) in 1980. We compare their interest vectors of their first year of publication and their last. Interest vectors $\mathbf{x}^{i,S}$ were built using the second level of the classification scheme with $69$ PACS.}
\label{fig:hist}
\end{figure}

\begin{figure}
\includegraphics[scale=0.3]{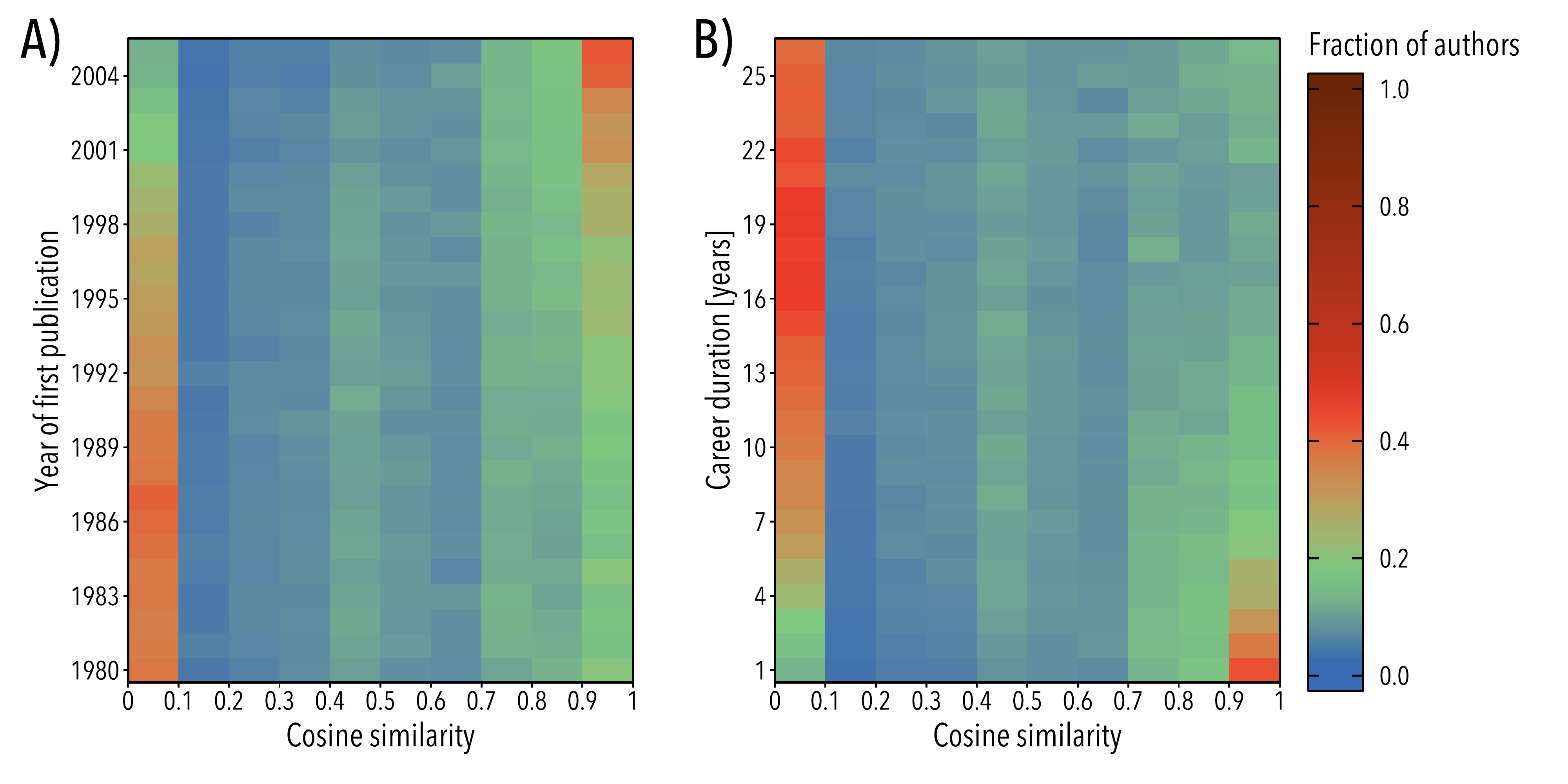}
\caption{Cosine similarity distribution of authors' interest between the first and last years of their careers measured using the second level of the classification scheme. On the left, figure A, distribution as a function of when they started their careers. On the right, figure B, distribution as a function of the duration of their careers. In both cases there is a clear tendency towards exploration.}
\label{fig:similarity}
\end{figure}

\begin{figure}
\includegraphics[scale=0.5]{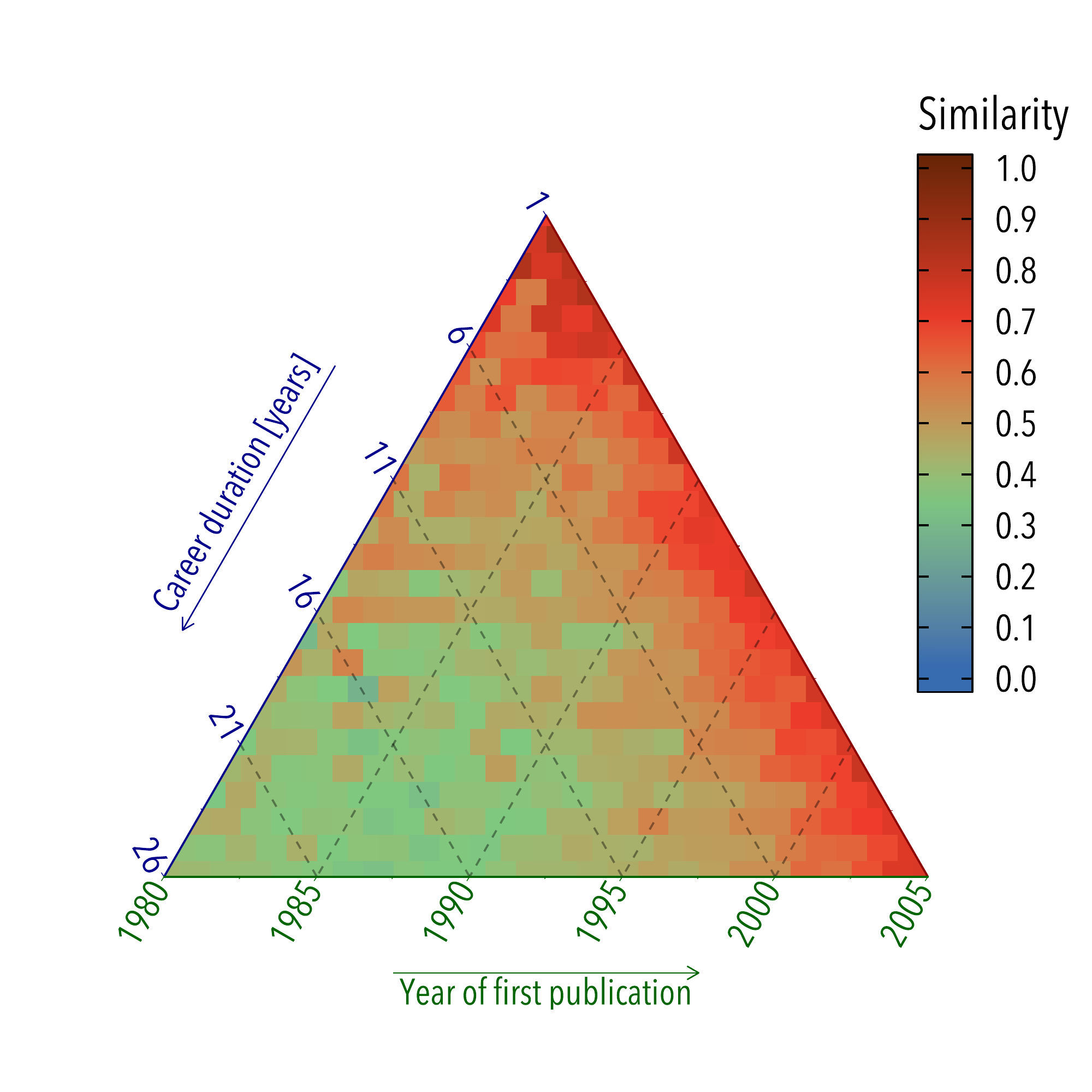}
\caption{Average similarity of authors' interest between the first and last years of their careers measured using the second level of the classification scheme. Regardless of the first year of publication the tendency towards exploration is higher the larger the duration of a career is.}
\label{fig:similarity_mean}
\end{figure}

\begin{figure}
\includegraphics[width=\linewidth]{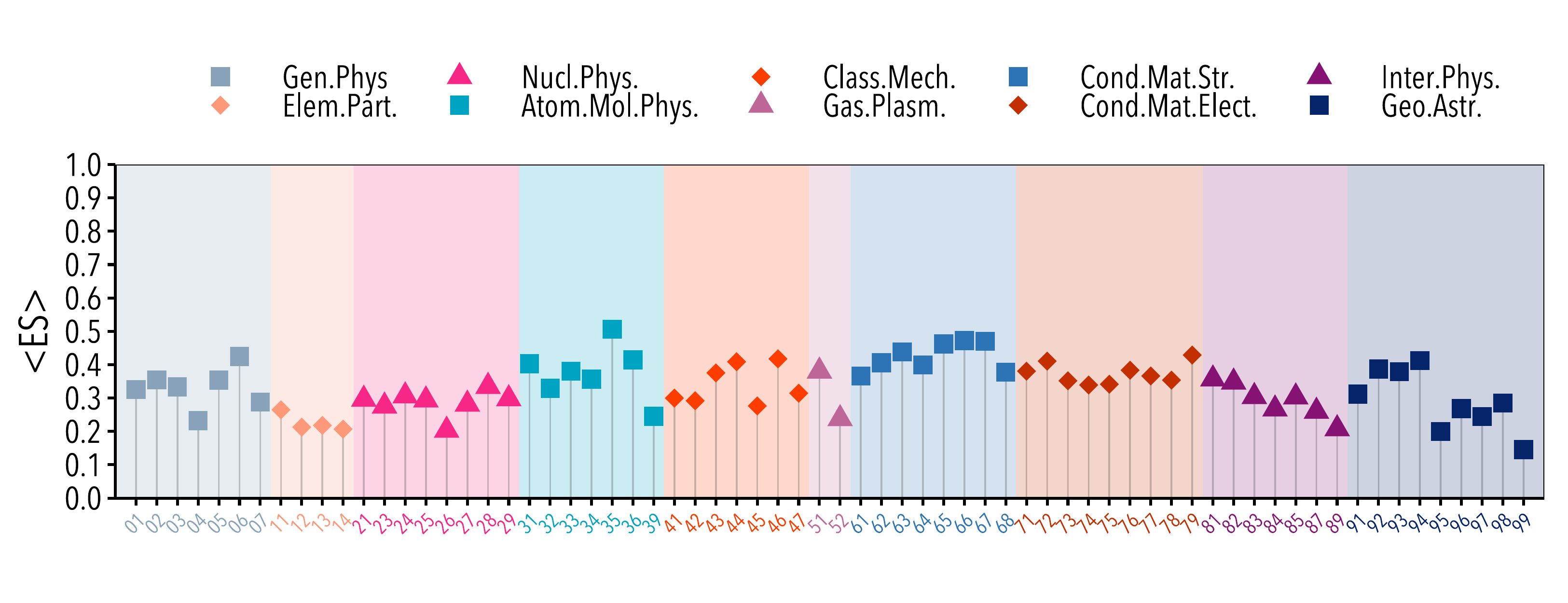}
\caption{Average exploration share (ES) as a function of the first topic used by each author.}
\label{fig:Exploration}
\end{figure}

How does the scientific interest of researchers change across time? To provide answers to this question let us first measure the similarity of scientific production at different careers stages. For simplicity, we consider the first ($f$) and the last ($l$) year of activity in our dataset. Then, for each career stage $S$, $S\in [f,l]$,  and author $i$ we build a vector $\mathbf{x}^{i,S}$ of size equal to the number of PACS at the classification level under consideration, i.e., $10$ at the first and $69$ at the second level, etc.  The vectors are constructed so that the generic component, $x^{i,S,}_{\alpha}$, describes the fraction between the number of times the PACS $\alpha$ has been used and the total number of PACS adopted. Thus, the components quantify the share of interest, in a specific year, towards the various PACS. In order to determine the similarity between vectors we use the cosine similarity, $\theta=\cos(\gamma)=\frac{\mathbf{A} \cdot \mathbf{B}}{||A||_2 ||B||_2}$, defined for each pair of vectors $\mathbf{A}$ and $\mathbf{B}$. To start getting a feeling about the distribution of the similarities, we first consider all authors that published their first papers in $1980$ and compare the first year of publication with their last, using the $69$ second level PACS. As it is clearly seen in Figure~\ref{fig:hist}, two tendencies are followed by the largest number of authors: $\theta >0.9$ and $\theta<0.1$. Thus, authors were more likely to keep working in the same topics potentially exploring few others, or instead change almost completely the subject of investigation. It is important to notice how the tendency towards a substantial change in research interests is embraced by a higher number of authors while the second, third and forth more likely values are concentred for high values of $\theta$ which describe authors covering similar topics during their career.\\
These first results demonstrate that exploration seems to be the preferred strategy. Does this apply also to authors that started their career in different years? Also, how does $\theta$ depend on the career duration? In Figure~\ref{fig:similarity} we answer to these questions. In particular, in Figure~\ref{fig:similarity}A we show the similarity as a function of the starting year for the second level PACS. Interestingly, we see a similar trend. Strong exploration (cosine similarity $<0.1$) seems to be the preferred strategy with strong exploitation (cosine similarity $>0.9$) the second most abundant trend. The only exception are younger scientists --who published their first paper in the $00s$-- that seem to prefer exploitation. The reason behind this result could be given by the fact that younger scientists are usually pursuing their mentors research line and have not outlined their own research agenda yet. Moreover, our dataset is limited to $2006$ thus for authors that started working in the early 2000s we have access to only the initial phase of their careers. To test this hypothesis, in Figure~\ref{fig:similarity}B we show the similarity as a function of the career duration. The plot shows an interesting trend. Short career durations (less than $4$ years) show a higher propensity to exploitation, while longer careers usually mean a tendency to exploration. This reinforces our idea that younger scientists tend to follow the research interests of their mentors and that the shift in the research line occurs after the Ph.D. --the crossover in Figure~\ref{fig:similarity}B takes place around 4 or 5 years of career, the usual duration of Ph.D. studies in many countries--. This finding is in line with the analyses done by Battiston et al~\cite{battiston2019taking} that showed how the average time of the first transition between fields is around $3-7$ years depending on the field. However, we also note that an alternative and plausible hypothesis is that this result reflects a change in the way science is done: the culture of ``publish or perish" indeed enforces incremental publications at the cost of undermining exploration or more risky career paths. In the future, when we will have more data about the evolution of younger authors, we would be in a better position to discriminate among these two scenarios.\\
As a way to consolidate all the previous observations, in Figure~\ref{fig:similarity_mean} we plot the average similarity as a function of the first year of publication and the career duration. Interestingly, we don't see any clear dependence on the starting year. The crucial difference is instead on the career duration. Indeed, the largest values of similarity are concentrated in the region of short careers. Authors with long careers instead are more prone to exploration.
By using the same vectors we can measure, more directly, the share of interest kept towards a set of PACS previously used (exploitation) and towards a set of new PACS (exploration). For each author we quantify the fraction of new and old PACS comparing the different career stages. In particular, we define the exploration share (ES) of author $i$ at stage $l$ or her career as:
\begin{equation}
ES_i^{l}=\sum_{\alpha}x^{i,l}_\alpha ( 1- H[x^{i,f}_\alpha])
\end{equation}
where $H[n]$ is a step function such that $H[n]=1$ for $n\ge0$. In words, $ES_i^l$ is the sum of the components of $x^{i,l}$ that were zero in $x^{i,f}$, thus the share of research activity towards new PACS. As vectors are normalised, the exploitation share is instead $1-ES_i^{l}$. By studying the exploration share of each author we can go a step further in our analysis and explore differences between different subfields. In Figure~\ref{fig:Exploration} we plot the average exploration value as a function of the first topic used by each author. In other words, we observe the tendency towards exploration differentiating between users starting in different fields and sub-fields. We note that Particle Physics, Nuclear Physics, Geology Astronomy and Astrophysics are less prone, on average, to explore different topics while the two Condensed Matter and Atomic and Molecular Physics are the ones with the highest exploration. We can speculate that this is due to the fact Particle, Nuclear and Astro Physics are very specialized and usually require large infrastructures while methods employed in other areas are more general. Looking inside each area we can see in some cases a large variability, e.g. in General Physics. Some sub-topics have a high ES like {\it Mathematical methods in Physics} (id. $02$) or {\it Metrology, measurements, and laboratory procedures} (id. $06$) while  {\it General relativity and gravitation} shows one of the lowest  propensity to exploration of the entire dataset. Along this line, an interesting example is topic id. $35$ {\it Experimentally derived information on atoms and molecules; instrumentation and techniques} that, despite a large proportion of papers (more than $800$), also presents the largest ES. This is probably due to the fact that PACS id. $35$ has been deleted from the $1995$ edition of the classification~\cite{pacs95} forcing all the scientists working on the topic to move to other PACS.

\begin{figure}
\includegraphics[scale=0.25]{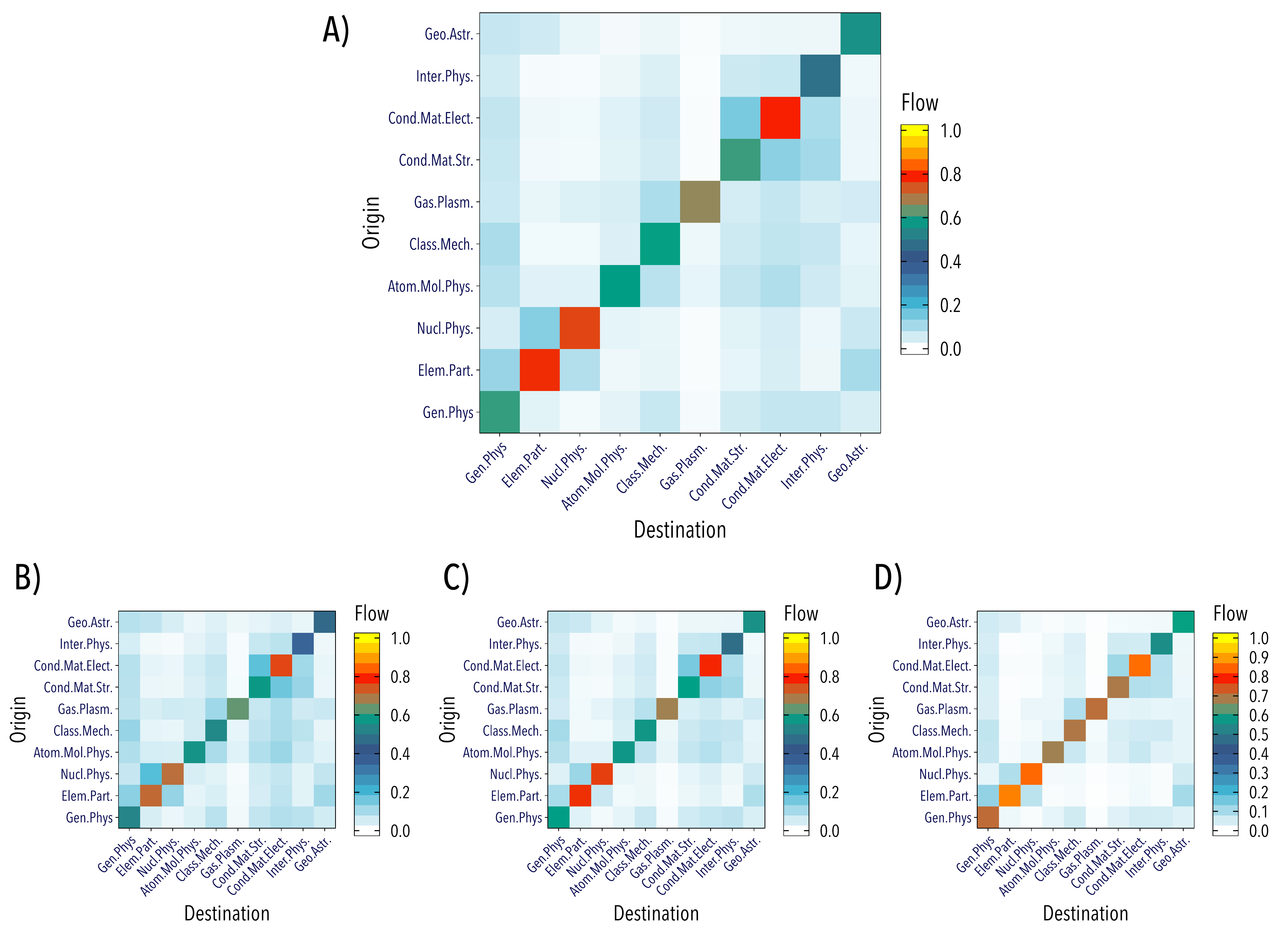}
\caption{Interest flow across decades: A, all decades; B, authors who started publishing in the 80's; C, authors who started publishing in the 90's; D, authors who started publishing in the 00's. Authors who start studying $y$ PACS end up sharing their interest across $x$ PACS. Each row is normalized over the number of authors who started using that PACS so that the diagonal represents the fraction of authors that kept some interest (see equation 2) in that PACS at the end of their career.}
\label{fig:matrices}
\end{figure}

So far we have quantified the tendency of authors towards exploration and exploitation. However, when authors explore new topics which ones do they consider? Are there exploration patterns more likely than others? How do these depend of the starting set of interests? To answer these questions, we first build origin-destination matrices by considering the flow of researchers from PACS to PACS comparing the first and last year of activity. Clearly, this analysis neglects trajectories between the two periods, but it offers a first indication of the general trends in scientific interest contrasting two distinct career phases. Let's define the flow from PACS $\alpha$ to PACS $\beta$ as:
\begin{equation}
M_{\alpha,\beta}=\sum_{i} \left ( H[x^{i,l}_\alpha]H[x^{i,f}_\beta]\delta_{\alpha,\beta}+ (1-\delta_{\alpha,\beta})\frac{H[x^{i,l}_\beta]H[x^{i,f}_\alpha](1-H[x^{i,f}_\beta])}{\sum_{\gamma}H[x_\gamma^{i,f}]} \right).
\end{equation} 
Each element of the matrix considers all the authors (thus the sum over $i$). Furthermore, we have two types of elements: inside and outside the diagonal. The first term contributes to the diagonal elements ($\delta_{\alpha,\beta}$ is the Kronecker delta) and it assumes a value of $1$ for all the authors that kept working on the PACS $\alpha$ in the first ($f$) and last ($l$) year of career. Thus, the term counts how many authors kept interest in the same PACS. The second term instead contributes to the off-diagonal elements. The numerator is equal to $1$ for all the $\alpha-\beta$ pairs that respect the following conditions: the author $i$ i) did not use $\beta$ in the first year, ii) used $\beta$ in the last year, iii) used $\alpha$ in the first year. The denominator instead is equal to the number of different PACS used in the first year. Thus, we connect each PACS used in the first year with those used only in the last year as a way to map the evolution in interest and a transition from a set of topics to another set. In Figure~\ref{fig:matrices} we report the results considering the first level of the classification. The first panel is obtained considering all the authors in the dataset. The other three instead are obtained distinguishing the researchers by the year of first activity.  Some important observations are in order. In general, the diagonal, for all the years, contains the largest values. This result, combined with Figs.~\ref{fig:hist},\ref{fig:similarity} and \ref{fig:similarity_mean}, highlights an interesting phenomenon. While most of the authors after 4 or 5 years of career almost totally change their interests, they usually remain in the larger area of Physics where they started. In a sense, in each author there is a strong tendency to explore but only within sight from their initial topic. This latter result is the empirical confirmation of the ``essential tension'' between risky and conservative strategies.

Looking at how physicists move outside their original area, other interesting trends emerge too. One of them is that the tendency towards exploitation is particular strong for scientists starting their career in Physics of Elementary Particles, Nuclear Physics, and Condensed Matter  (Electronic Structure, Electrical, Magnetic, and Optical Properties) while another interesting observation concerns the sub-field of Physics of Gases, Plasmas and Electric Discharges (id $5$). Indeed, across years we can observe that, with respect to all the other topics, this is the one that is less likely to ``attract" researchers from other areas. A similar result holds, although more nuanced, for the field of Geophysics, Astrophysics, and Astronomy. On the other hand, as far as exploration is concerned, the field that is able to attract more authors that initiated their publication record in other subjects is General Physics, which is by construction one of the most interdisciplinary fields. Moreover, from the matrices two clusters are clearly visible. The first is formed by Particle and Nuclear Physics. The second instead is formed by the two fields of Condensed Matter and Interdisciplinary Physics. The presence of such cluster implies that, for example, authors starting in Particle Physics are more likely, in case they explore new topics, to move towards  Nuclear Physics. Finally, it is interesting to note how these patterns are preserved across different generations of researchers that started publishing in different decades.

\begin{figure}
\includegraphics[width=\columnwidth]{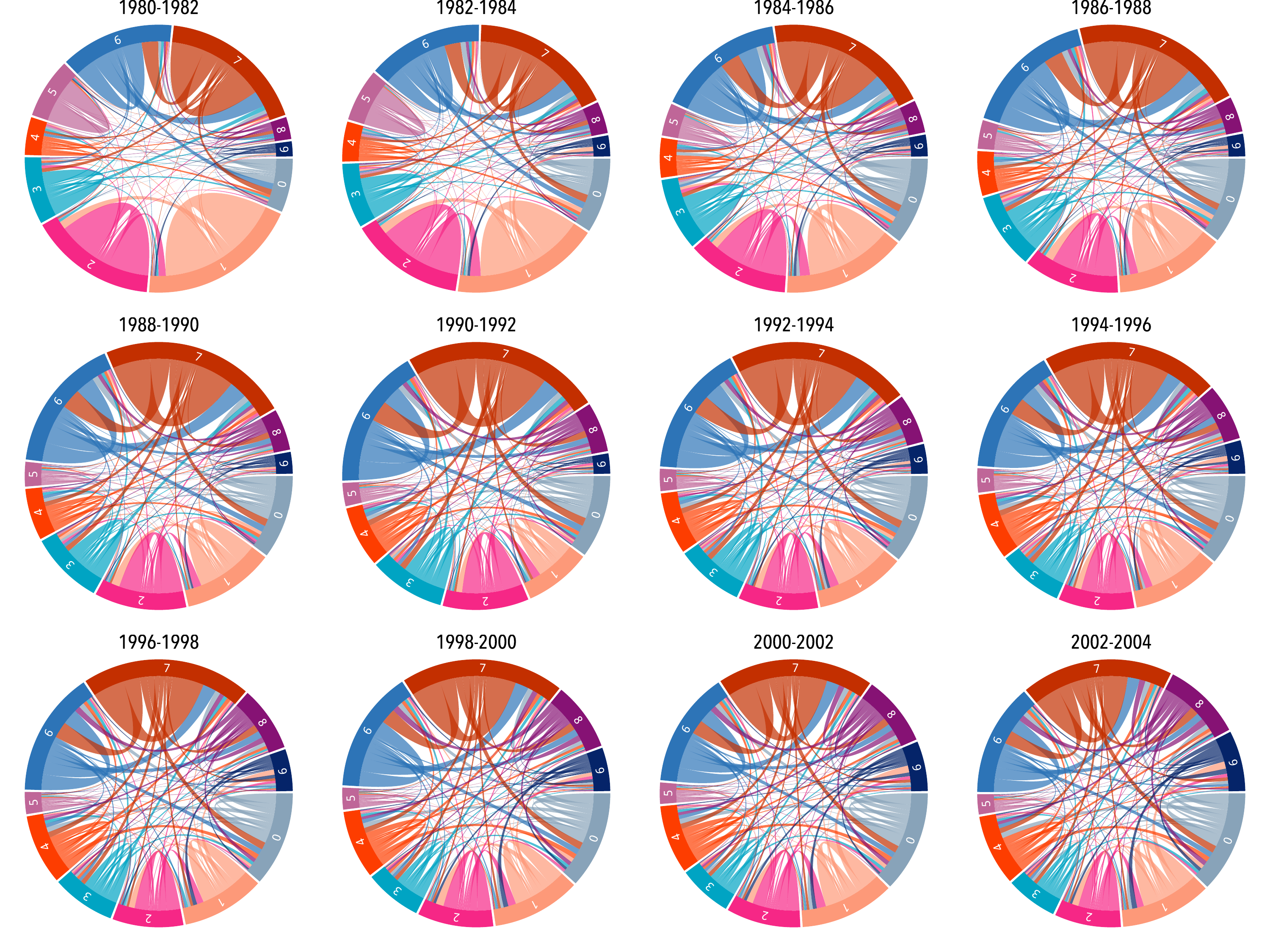}
\caption{Authors flow across fields. For each time window (2 years), PACS are disposed in a circle. The width of each arc is proportional to the number of authors who published at least one paper in that field (and that published at least another one the year after). Each field is represented by its PACS number, see table \ref{table:1}. Links represent the number of authors who, having published at least one paper in the source field on the right, also publish at least one paper in the target field on the left (at seen from outside the circle). Note that an author can publish papers in several fields in a given time window so that a link does not mean that she changed her field but that, at least, she has some interest on it. The plots are made using \emph{circlize} \cite{circlize}.}
\label{fig:figx}
\end{figure}

Overall, the results showed so far can be summarised as follows: i) even if exploration is the preferred strategy, usually it is confined within the first level of the classification, probably offering the right mix between exploration and exploitation, ii) exploration outside the first level is not random as the transition from some fields to others is more likely. These observations are in line with previous work done with different measures and metrics~\cite{pan2012evolution,battiston2019taking}. However, they are in contrast with the work done by Foster et al~\cite{foster2015tradition} and Jia et al~\cite{jia2017quantifying}. The first group focused on a different research area (Biomedical Chemistry) and studied $133$ awardees of scientific prizes. In that field, scientists seem to prefer exploitation than exploration. This opposite trend highlights how the essential tension might be a function of the area of study. The second group studied, as we do here, the APS dataset. However, they considered a subset of authors that published at least $16$ papers (their results do not change considering $12$ or $20$). Furthermore, they considered event time rather than real time (i.e. years). Thus the sequence of publication of each authors does not have gaps (years of inactivity are not accounted for). While this approach is quite useful to eliminate possible issues associated to burstiness, it mixes individuals with very different publication rates and at different career stages. The last point is particularly relevant as the scientific maturity and independence, often necessary for exploration, are not necessarily a function of the number of papers published (especially in some disciplines that feature large collaborations). Indeed, our results, as well as those by Battiston et al~\cite{battiston2019taking}, show that periods before and after the typical PhD duration ($3-7$ years) are characterized by very different tendencies toward exploration. The contrast between the two results highlights a very important point: the \emph{inclusion} principle used to select the sample of scientists under study, and the approach used to account for time, might influence the results. It is important to notice how each methodology features different pros/cons and effectively select a different sample (with possible overlaps). Cleary, more work needs to be done to explore the effects of different approaches aimed at define which publication record should be considered as signature of a professional scientist.\\
Up to now we have mapped the transitions, that is flows between topics, comparing the first and last year of activity in our database. Next, we deepen our investigation by mapping the flows as a function of time. To this end, we consider all authors that published a paper in year $t$ and/or $t+1$. Note that we adopted a two years window to increase the statistics. Then, we consider the fraction of such authors that published a paper also in year $t+2$ and/or $t+3$. For each bi-annual time window, we dispose PACS in a circle and connect them with links proportionally to how many authors used PACS $\alpha$ and then PACS $\beta$. The width of each arc is a measure of popularity. Aggregating the normalised flows built considering each author, we obtain a systemic map of the dynamics of interests between fields. In Figure~\ref{fig:figx} we show the results considering only the first level of the classification. Several observations are in order. During the first years we don't see much flow between fields. The authors that published in contiguous time windows did not change topics as much as in later times. In the period $1984-1986$, instead, we start seeing an increase in connectivity between fields signaling either the publication of multidisciplinary papers (articles containing PACS from different fields) and/or authors exploring different fields. In particular, we see a more intense mixing between the two different branches of Condensed Matter (6 and 7) as well as an out flow from General Physics (0) towards these two fields. The same patterns are repeated from $1986$ to $1988$. Note also that since the first time window, there is some mixing between Elementary Particle and Nuclear Physics (1 and 2). However, from $1990-1992$ on, the mixing between fields become more evident. Condensed Matter: Electronic Structure, Electrical, Magnetic, and Optical Properties (number 7), is the field with the largest out-flow towards others. Furthermore, interestingly enough, we observe a the raise in popularity of General, Interdisciplinary, and Astro Physics. Such increase is balanced by a decrease in popularity of Physics of Gases and Plasmas, Elementary Particles and Nuclear Physics. It is important to note that, by definition, the popularity is not a single measure of the number of papers written each year in each field. Indeed, it is modulated by the number of authors that wrote papers in two consecutive years. Our results are in line with the Physics ``census" recently conducted by Battiston et al~\cite{battiston2019taking} with a much larger sample of publication venues. We also mention that our dataset does not allow us to see later trends that Battiston et al~\cite{battiston2019taking} observed, such as spikes of productivity in $2010$ in Elementary Particle Physics or the relative reduction of Condense Matter in the last years.

\section{Conclusions}
In this work, we have analysed the different strategies adopted by researchers, during their career in the Physics community, and test the presence of  ``the essential tension'' between exploration and exploitation described by Kuhn~\cite{kuhn1979essential}. To do so, we mapped the evolution of interests in Physics in the last $30$ years relying on a dataset containing all the papers published in the APS journals in the period $1980-2006$. Defining a set of individual and global metrics we quantified the change in the PACS used by authors along their careers. Furthermore, we analyzed the source-destination matrices of authors and the network flows between different topics. We were able to detect which areas of Physics serve as ``donors'' of scientists to other areas and  which ones are more likely to ``receive'' a researcher. 

Even if our analysis has several limitations $-$e.g., our dataset is limited to year $2006$ and do not cover Physics papers published in multidisciplinary journals$-$, we indeed confirm the existence of such ``tension'' between exploring new fields and exploiting the knowledge acquired during previous years. Our results demonstrate that, even if the vast majority of the authors almost completely change their research interests during their career, they remain in the broader area of Physics --i.e. the first level of the PACS classification-- where they started. This ``explore with caution'' strategy seems to be the best tradeoff between the risk of moving to new fields and taking advance of the work done in the past. These findings are in line with, and complement, previous research that focused on Physics as scientific area. In fact, Jia et al~\cite{jia2017quantifying} have clearly identified subject proximity as a critical factor influencing authors' production. Pan et al~\cite{pan2012evolution} have shown how the networks constructed by using the co-occurrence between PACS densify in time and that such increase in connectivity is hierarchical: close sub-fields connect first. Our results, together with the work by Jia et al~\cite{jia2017quantifying}, suggest that such temporal dynamics might be indeed driven by the essential tension between exploration and exploitation faced by each author. It is important to notice however how our results are opposite to those presented by Foster et al~\cite{foster2015tradition}. As mentioned in the introduction, these authors found that in the area of Biomedical Chemistry exploitation is instead the preferred strategy. This contrasts with what we found in Physics, and raises an important question for future research: how does the essential tension affect different scientific areas? As mentioned above, our results are also opposite to the findings (in terms of the tendency towards exploration) of Jia et al~\cite{jia2017quantifying}. Despite that we used the same dataset, we adopted a very different inclusion principle (to select the sample of authors to study) and measured the career duration not in terms of papers published but in years. This raises another important question for future work: what constitutes a professional scientist and how should we study her career progression? Indeed, the literature is quite divided in this point. Battiston et al~\cite{battiston2019taking} for example considered only authors that published at least five papers. Jia et al~\cite{jia2017quantifying} studied only authors that published at least $16$ articles and Pan et al~\cite{pan2012evolution} did not impose any restrictions (although they did not focus on the evolution of single authors but rather on the evolution of disciplines).

Another interesting result stemming from our analysis is that the tendency towards exploration is more marked for scientists with longer careers, with a minimum of $4$ or $5$ years to start exploring. While this minimum value is probably related to the length of Ph.D. studies, it also highlights that, unlike exploitation, exploration requires longer time to payback. This conclusion is in line with the work by Battiston et al~\cite{battiston2019taking} who, with different metrics, have shown that the average time for the first transition between fields to take place, is within $3-7$ years, depending on the starting area. Additionally, by defining the ``migration flows'' of authors between topics, we identified the areas of Physics with the larger vocation to explore and the most probable paths for scientists leaving each area. Physics of Elementary Particles and Nuclear Physics turned out to be the areas with the lowest tendency for exploration but, interestingly,  they form a closed cluster with an almost balanced interchange of scientists --probably due to the relatedness of topics and methodology used--. Another tight cluster is the one including the two Condensed Matter and Interdisciplinary Physics. In this case Cond. Mat.  (Electronic Structure, Electrical, Magnetic, and Optical Properties) is also a very closed area but with a steady flow of researchers from and to the other two areas. Interestingly, these findings are in line with the work by Battiston et al~\cite{battiston2019taking} that, however, studied a much larger set of Physics journals and papers well beyond those published by the APS.  

In a nutshell, our results, even if largely in line with previous research, depict a more nuanced portrait of the evolution of research interests than previously thought~\cite{jia2017quantifying,pan2012evolution,foster2015tradition,battiston2019taking}. Taking into account the first and second levels of the PACS classification we demonstrated that physicists indeed explore during their career but only in the proximity of their initial research topic. In some sense we can say that the area of the first year of a researcher marks the rest of her career but that inside each area there is ample space to explore new interests. Taken together, our results highlight the high dynamism of the Physics community and the lines of evolution of the field. Finally, we believe that the results presented in this work can help the design of specific policies to foster the future advancement of Physics and related scientific disciplines. 

%{\color{red}Note: for all plots I've used ggplot2 \cite{ggplot2}, ggtern \cite{ggtern} for the 3rd and circlize \cite{circlize} for the last one. I don't know where should we add the references, maybe in a materials and methods section?}

\section{Acknowledgements}
This material is based upon work supported by, or in part by, the U. S. Army Research Laboratory and the U. S. Army Research Office under contract/grant number W911NF-18-1-0376. A.A. acknowledges support from Santander via the ``Universities International Mobility Awards`` program and of the FPI doctoral fellowship program from MINECO (Spain). S.M. acknowledges support from the Spanish State Research Agency, through the Mar\'ia de Maeztu Program for Units of Excellence in R\&D (MDM-2017-0711 to the IFISC Institute). Y.M. acknowledges partial support from the Government of Arag\'on, Spain through a grant to the group FENOL (E36-17R), by MINECO and FEDER funds (grant FIS2017-87519-P) and by Intesa Sanpaolo Innovation Center. The funders had no role in study design, data collection and analysis, decision to publish, or preparation of the manuscript. 

\section{Authors Contributions}
Designed the research: SM, NP, YM. Performed the analysis: AL. Analyzed the results: AL, SM, NP, YM. Wrote the manuscript: AL, SM, NP, YM. All authors read and approved the final version of the manuscript.

\section{Conflict of Interests}
All the authors declare no conflict of interests.

%\bibliography{refs}

\end{document}